\documentclass[aps,prd,twocolumn,amsmath]{revtex4-1}
\usepackage{times,amsbsy,amsmath,amsfonts,graphicx,float}
\usepackage{color,morefloats,rotating,srcltx,slashed}
\usepackage{multirow,bm,verbatim,tabularx,bbding,threeparttable}
\definecolor{dblue}{rgb}{0.00,0.00,0.75}
\usepackage[colorlinks,urlcolor=dblue,linkcolor=dblue,citecolor=dblue]{hyperref} 
\usepackage{setspace}
\allowdisplaybreaks[4]

\begin{document} 

                      \title{A new look at the $P_{cs}$ states from a molecular perspective} 
                                     
\author{Albert Feijoo$^1$}	          \email{edfeijoo@ific.uv.es}
\author{Wen-Fei Wang$^{1,2}$}	  \email{wfwang@ific.uv.es}
\author{Chu-Wen Xiao$^3$}  	         \email{xiaochw@csu.edu.cn}
\author{Jia-Jun Wu$^4$}  	         \email{wujiajun@ucas.ac.cn}		
\author{Eulogio Oset$^1$}	         \email{oset@ific.uv.es}
\author{Juan Nieves$^1$}	         \email{jmnieves@ific.uv.es}
\author{Bing-Song Zou$^{4,5}$}      \email{zoubs@itp.ac.cn}
		
	\affiliation{$^1$Departamento de F\'{\i}sica Te\'orica and IFIC, Centro Mixto Universidad de
	                 Valencia-CSIC Institutos de Investigaci\'on de Paterna, Aptdo.~22085, 46071 Valencia, Spain\\           
	                 $^2$Institute of Theoretical Physics, Shanxi University, Taiyuan, Shanxi 030006, China\\      
	                 $^3$School of Physics and Electronics, Central South University, Changsha 410083, China\\ 
                         $^4$School of Physical Sciences, University of Chinese Academy of Sciences (UCAS), Beijing 100049, China\\
                         $^5$Key Laboratory of Theoretical Physics, Institute of Theoretical Physics, Chinese Academy of Sciences, 
                                 Beijing 100190, China}	

\date{\today}

%XXXXXXXXXXXXXXXXXXXXXXXXXXXXXXXXXX%
\begin{abstract}
We have a look at the $P_{cs}$ states generated from the interaction of $\bar D^{(*)} \Xi^{(\prime*)}_c$ coupled channels. 
We consider the blocks of pseudoscalar-baryon $({\frac12}^+, {\frac32}^+)$ and vector-baryon $({\frac12}^+, {\frac32}^+)$, 
and find $10$ resonant states coupling mostly to $\bar D \Xi_c, \bar D^* \Xi_c,\bar D \Xi'_c, \bar D^* \Xi'_c,\bar D \Xi^*_c$ and 
$\bar D^* \Xi^*_c$. A novel aspect of the work is the realization that the $\bar D \Xi_c,\bar D_s\Lambda_c$  or 
$\bar D^* \Xi_c,\bar D^*_s\Lambda_c$ channels, with a strong transition potential, collaborate to produce a larger attraction 
than the corresponding states $\bar D \Sigma_c,\bar D\Lambda_c$ or $\bar D^* \Sigma_c,\bar D^*\Lambda_c$ appearing in the 
generation of the strangenessless $P_{c}$ states, since in the latter case the transition potential between those channels is zero. The extra attraction obtained in the $\bar D \Xi_c,\bar D^* \Xi_c$ pairs preclude the association of these channels to the  $P_{cs}(4338)$ 
and $P_{cs}(4459)$ states respectively. Then we find a natural association of the $P_{cs}(4338)$ state coupling mostly to 
$\bar D^* \Xi_c$ while the $P_{cs}(4459)$ is associated to the state found that couples mostly to $\bar D \Xi'_c$.  Four more 
states appear, like in other molecular pictures, and some of the states are degenerate in spin. Counting different spin states we 
find $10$ states, which we hope can be observed in the near future.
\end{abstract}

%\pacs{13.20.He, 13.25.Hw, 13.30.Eg}
\maketitle
%XXXXXXXXXXXXXXXXXXXXXXXXXXXXXXXXXX% @ Begin

%%%%%%%%%%%%%%%%%%%%%%%%%%%%
\section{Introduction}\label{sec:1}%%%========SEC1

The discovery of the two pentaquark states by the LHCb collaboration in $2015$ in the $J/\psi p$ spectrum of the $\Lambda^0_b \to J/\psi K^- p$ decay \cite{LHCb:2015yax,LHCb:2015qvk} was a turning point in hadron physics which stimulated a tremendous amount of theoretical work aimed at describing the nature of such states. This work is reflected in a large number of review papers \cite{Chen:2016qju, Hosaka:2016pey,Chen:2016spr,Lebed:2016hpi,Esposito:2016noz,Oset:2016lyh,Guo:2017jvc,Ali:2017jda,Olsen:2017bmm,Karliner:2017qhf,Yuan:2018inv,Liu:2019zoy} and different pictures have been proposed to explain these states, in most cases as molecular states of meson baryon or as compact pentaquark states. In $2019$ the LHCb collaboration updated the former results with additional  data from Run-2, where three narrow structures were observed \cite{LHCb:2019kea}, the $P_c(4312)$, $P_c(4440)$ and $P_c(4457)$. Interestingly, in Refs.~\cite{Wu:2010jy,Wu:2010vk} predictions were done in $2010$ about such states of molecular nature, which agreed qualitatively with the states found. The work of \cite{Wu:2010jy,Wu:2010vk} was revisited from the point of view of heavy quark spin symmetry \cite{Isgur:1989vq,Neubert:1993mb,Manohar:2000dt} in \cite{Xiao:2013yca} and the only parameter of the theoretical model, the regulator of the meson-baryon loops, was fine tuned to get a good description of the experimental data. The same work in Refs.~\cite{Wu:2010jy,Wu:2010vk} predicted also states of hidden charm and strange content, $c\bar csqq$, and work followed in Refs.~\cite{Santopinto:2016pkp,Chen:2016ryt,Wang:2019nvm}. One of such states, named $P_{cs}(4459)$, was found again by the LHCb collaboration in the $J/\psi \Lambda$ invariant mass distribution of the $\Xi^-_b \to J/\psi K^- \Lambda$ \cite{LHCb:2020jpq}, which had been suggested in \cite{Santopinto:2016pkp,Xiao:2019gjd,Chen:2015sxa} as an ideal reaction to look for these states. The experimental finding drew much theoretical attention intended to explain the nature of the state found, while making further predictions \cite{Chen:2020uif,Wang:2020eep,Azizi:2021utt,Ozdem:2021ugy,Peng:2020hql,Chen:2020kco,Chen:2021tip,Liu:2020hcv,Zhu:2021lhd,Liu:2020ajv,Wang:2022mxy,Ozdem:2022kei}. Furthermore, very recently, a new $P_{cs}$ state, $P_{cs}(4338)$ was found in the $J/\psi \Lambda$ mass distribution of $B^- \to J/\psi \Lambda \bar p$ decay  \cite{LHCb:2022jad}. 

After the finding of the $P_c$ states, predictions for $P_{cs}$ states were made in Ref.~\cite{Xiao:2019gjd} within the heavy quark spin symmetry scheme used successfully in Ref.~\cite{Xiao:2013yca}. The predictions made could be considered in rough qualitative agreement with the experimental states found later. Yet, with the perspective given by the two states found and the theoretical work done after the observation of the latest $P_{cs}(4338)$ state, a fresh look at the problem is most opportune and we address this issue here.

In \cite{Wang:2022neq}, the QCD sum rule method was used and the molecular structure of $\bar D \Xi_c,\bar D^* \Xi_c$ for the $P_{cs}(4338)$ and $P_{cs}(4459)$ states was suggested. In \cite{Meng:2022wgl} the boson exchange model was used and, once again the  $\bar D \Xi_c$ is associated with the $P_{cs}(4338)$ while the $\bar D^* \Xi_c$ , with two structures as ${\frac12}^-, {\frac32}^-$, is associated to the $P_{cs}(4459)$. At the same time predictions are made of states of $\bar D^{(*)} \Xi'_c$ structure. The lineshape of the $J/\psi \Lambda$ mass distribution is considered in \cite{Meng:2022wgl} assuming that the $P_{cs}(4338)$ state is made from the $\bar D^0 \Xi_c^0$ and $D^-\Xi_c^+$ components. A different view is presented in \cite{Burns:2022uha} suggesting that the $P_{cs}(4338)$ peak, together with another peak, are the result of triangle singularities, relying on a formalism that contains a few free parameters. A constituent quark model is investigated in \cite{Ortega:2022uyu} looking at the molecular meson-baryon structures and 9 states are obtained of $\eta \Lambda, J/\psi \Lambda, \bar D^{(*)}_s \Lambda_c,\bar D^{(*)}\Xi_c, \bar D^{(*)}\Xi'_c,\bar D^{(*)}\Xi_c^{(*)}$ type. Once again the $P_{cs}(4338)$ is associated mostly to the $\bar D \Xi_c$ state, although the $\bar D^{(*)}_s \Lambda_c$ channel also has a sizable component, while the $\bar D^{*}\Xi_c^{*}$ has a small one. In \cite{Wang:2022mxy} the molecular picture is invoked suggesting that the $P_{cs}(4459)$ may contain two substructures of $\bar D^{*} \Xi_c$ with $J^P={1/2}^-,{3/2}^-$ and there is the corresponding $\bar D \Xi_c$ state with $J^P={1/2}^-$.  A pentaquark structure is assumed for the two states in \cite{Ozdem:2022kei} and magnetic moments are evaluated using QCD light cone sum rules. In \cite{Zhu:2022wpi}, a potential kernel constructed from meson exchange is used in the Bethe-Salpeter equation and association of  $\bar D \Xi_c$ to the $P_{cs}(4338)$ is made while $\bar D^{*} \Xi_c$ appears in ${1/2}^-,{3/2}^-$, mostly in $J^P={1/2}^-$, and is associated to the  $P_{cs}(4459)$. In \cite{Nakamura:2022jpd}, the authors determine the pole of $P_{cs}(4338)$ within a coupled channel framework by fitting the various distributions of the final states in $B^-\to J/\psi \Lambda \bar{p}$ and also confirm that the peak in the spectrum of $J/\psi \Lambda$ cannot come just from a cusp. In that model, the $P_{cs}(4338)$ is also related to the $\bar{D}\Xi_c$ channel and there is another possible pole named as $P_{cs}(4255)$ hinted by one datum around the $\bar{D}_s \Lambda_c$ threshold  which is, however, compatible with a statistical fluctuation \footnote{Tim Gerson private communication.}. In \cite{Chen:2022wkh}, the meson-baryon interaction is parametrized according to $SU(3)_f$ light flavor and heavy quark symmetries, and coupled channels are considered. Once again the $P_{cs}(4338)$ state is associated to the $\bar D \Xi_c$ component, while two states of $\bar D^* \Xi_c$ nature with $J^P={1/2}^-,{3/2}^-$ would be associated to the $P_{cs}(4459)$, although a state of $\bar D \Xi'_c$ nature appears around $4434$~MeV with a very small width, which would make it incompatible with the observed $P_{cs}(4459)$ state. One interesting feature of \cite{Chen:2022wkh} is the important effect of the $\bar D \Xi_c - \bar D^{(*)}_s \Lambda_c$ coupled channels that reduces the binding of the $\bar D \Xi_c$ state considered as single channel. We shall discuss the effect of coupled channels in detail in this paper. In \cite{Yang:2022ezl}, using as the source of interaction $\sigma, \rho, \omega$ exchange, once again, the association of the $P_{cs}(4338)$ to $\bar D \Xi_c$ and $P_{cs}(4459)$ to $\bar D^* \Xi_c$ is favored.

One important issue affecting both the $P_{c}$ and $P_{cs}$ states is the present status concerning the spin-parity of the states. While all the theoretical studies suggest $J^P$ values, mostly $J^P={1/2}^-,{3/2}^-$, for the different states found, experimentally these quantum numbers are not yet established. It is fair to assume that future runs of the LHCb with more statistics will allow the determination of these important quantities, which will help put some order into the different theoretical predictions. One promising path of progress in this direction is given by the recent work \cite{Zhang:2023czx} where the spectrum, analyzed with the help of machine learning techniques, sheds light on this particular issue. 

The present work offers and explanation of the two $P_{cs}$ pentaquark states in a molecular picture which successfully reproduced the observed $P_c$ states. There are however alternative proposals based on compact pentaquark states starting with the interaction of the quarks themselves. There are multiple works on the issue which is not our purpose to discuss here, but it becomes instructive to discuss some of the main differences between these pictures and how future experiments can help to find the picture adjusting more closely to the reality. Reviews on the subject and comparison between models have been quoted before, and adjusting more to the compact pentaquark structures one can mention \cite{Lebed:2016hpi,Esposito:2016noz,Ali:2017jda,Liu:2019zoy}. One characteristics of these models is the large freedom to make substructures of quarks. In contrast to the molecules that have the quarks clustered into a meson and a baryon, the compact pentaquarks rely on diquark structures. Yet, even then, there is a large freedom on how to clusterize the five quarks. As an example we choose a few works that evidence this freedom. In \cite{Ali:2019clg,Ali:2016dkf,Ali:2017ebb} compact hidden-charm diquark-diquark-antiquark pentaquarks structures are considered, while in \cite{Giron:2021sla,Giron:2021fnl} the authors consider diquark-triquark structures. In \cite{Shi:2021wyt} two diquarks and an antiquark structures are taken, as in \cite{Ali:2019clg,Ali:2016dkf,Ali:2017ebb}, but there are differences in the details of the interaction, which lead to large differences in the obtained spectra. Even between \cite{Ali:2019clg} and \cite{Ali:2016dkf,Ali:2017ebb} by the same group, differences of mass of $100$-$300$~MeV are found for the same states. There are also differences in the assignment of spin and parity to the observed states. Indeed, while in \cite{Shi:2021wyt} states of negative parity are preferred, in \cite{Ali:2019clg} some states would be of positive parity and others of negative parity. In \cite{Giron:2021sla,Giron:2021fnl} structures corresponding to orbital excitation of the clusters with $L=1$, are preferred, leading to positive parity states. One common characteristics of all these models is the large amount of states predicted, of the order of $30$ in \cite{Ali:2019clg,Shi:2021wyt}, but it is discussed in \cite{Ali:2019clg} that if arguments of heavy quark spin symmetry are used, the number is reduced. Another problem related to these models is the width of the states which is only discussed qualitatively but no precise numbers are given. The widths appear naturally large and arguments are given some times to justify that restrictions imposed by heavy quark symmetry could be responsible for the small observed widths \cite{Ali:2019clg}. Different arguments are given in \cite{Brodsky:2014xia}, suggesting that the small widths originate from the significant spatial separation between the diquark or triquark quasiparticles, which hinders the rearrangement of their component quarks into color-singlet hadrons. On the other hand in \cite{Giron:2021sla} the large widths expected from states with orbital angular momentum $L=0$ are taken as an argument to prefer the $L=1$ pentaquarks, that lead automatically to positive parity. 
  
In the molecular picture there is less freedom and one obtains fewer states. The states obtained are all of negative parity since the S-wave attractive interaction is applied between negative parity meson and a positive parity baryon for $P_{c}$ and $P_{cs}$ states. The widths obtained are small and they result automatically from decay to states of meson baryon type, used in the same set of the coupled channels which are allowed to interact in a coupled channels unitary scheme. 
    
Concerning the $P_{cs}$ states, the works discussed above only refer to the $P_{cs}(4459)$ and the new $P_{cs}(4338)$ is not discussed. The width of the $P_{cs}(4459)$ state is not addressed either. It would be interesting to extend the predictions of these models to the new state and its properties. In any case, the discussion done above leads us to the conclusion that to make advances in the field and pin down the nature of the $P_{c}$ and $P_{cs}$ states, the experimental determination of the spin and parity of these states is crucial. On the other hand, both the compact pentaquark models and the molecular ones predict more states than have been observed. Since all the models, adjusting their input to present states, make different predictions for other states, the likely observation of new states in future runs of present facilities, with more statistics, will also provide very valuable information to help us get a closer look at the structure of these states. 

%%%%%%%%%%%%%%%%%%%%%%%%%%%%
\section{Formalism}\label{sec:2}%%%========SEC2 

We start from the formalism of \cite{Xiao:2019gjd}, which used heavy quark spin symmetry (HQSS) together with the local hidden 
gauge approach (LHG) \cite{Bando:1984ej,Bando:1987br,Meissner:1987ge,Nagahiro:2008cv} extrapolated to the charm sector in 
order to determine the coefficients of the lowest order (LO) HQSS effective theory scheme.  
We divert a bit here from this picture and evaluate all the matrix elements of the coupled channels potential using the local
hidden gauge approach. There are two reasons for it. The extrapolation of the local hidden gauge approach to the heavy quark 
sector satisfies strictly the rules of HQSS for the dominant terms. Indeed, this approach relies upon the 
exchange of vector mesons. The leading order dominant terms correspond to the exchange of light vectors, where the heavy quarks are 
spectators, which automatically makes the matrix elements independent of the heavy quarks and the rules of heavy quark 
symmetry are automatically fulfilled. However the LHG approach also allows for the exchange of heavy vector mesons which are
suppressed by a $(m_V/m_{D^*})^2$ factor, and hence, vanishing in the infinite heavy quark mass limit, but still non negligible 
at the charm quark mass scale.  Second, in \cite{Xiao:2019gjd} the HQSS rules were extended to some terms 
which are subdominant, for instance to relate some non diagonal 
pseudoscalar-baryon (PB) and vector-baryon (VB) components. While this had minor numerical effects, it is not justified from
the fundamental point of view and the use of the strict LHG approach avoids this. Another novelty in the present approach is the use of the cutoff regularization of the loop functions $G_l$ instead of the dimensional regularization employed in \cite{Xiao:2019gjd}. The reason is that in the charm sector the $G$ functions in dimensional regularization can become positive below threshold, eventually allowing the appearance of bound states with a repulsive potential \cite{Wu:2010rv}. On the other hand we rely on the fact that the PB and VB components are related by the exchange of pseudoscalar mesons and their mixing is small as discussed in detail in
 Appendix B of Ref.~\cite{Dias:2021upl}. This said, we have now four block of coupled channels that we do not mix:
\begin{itemize}
  \item [1)] $PB (\frac12^+)$: Pseudoscalar meson-baryon of $J^P=\frac12^+$, 
                 $\eta_c\Lambda(4099.6)$, $\bar{D}_s\Lambda_c(4259.8)$, $\bar{D}\Xi_c(4336.3)$, 
                 $\bar{D}\Xi_c^\prime(4445.7)$;
  \item [2)] $PB (\frac32^+)$: Pseudoscalar meson-baryon of $J^P=\frac32^+$,
                 $\bar{D}\Xi_c^*(4512.9)$; 
  \item [3)] $VB (\frac12^+)$: Vector meson-baryon of $J^P=\frac12^+$, 
                 $J/\psi\Lambda(4212.6)$, $\bar{D}^*_s\Lambda_c(4398.7)$, $\bar{D}^*\Xi_c(4477.6)$, 
                 $\bar{D}^*\Xi_c^\prime(4587.0)$;  
  \item [4)] $VB (\frac32^+)$: Vector meson-baryon of $J^P=\frac32^+$,
                 $\bar{D}^*\Xi_c^*(4654.2)$;                        
\end{itemize}
where the numbers in parenthesis are the threshold masses of the coupled channels in units of MeV. In the nomenclature $PB(1/2^+)$, 
etc., the $1/2^+$ or $3/2^+$ refers to the $J^P$ of the baryon alone.

The wave functions of the elementary baryons in terms of quarks are taken as in \cite{Xiao:2019gjd} (section 2.3), where, following \cite{Roberts:2007ni},
we isolate the heavy quark and put the spin-flavor symmetry of the wave function in the light quarks. This allows for an easy
evaluation of the $VBB$ vertices in terms quarks, while the $PPV$ of $VVV$ vertices are trivial since mesons are simple $q\bar{q}$
structures. Then we obtain the transition potentials between channels as
\begin{eqnarray}
  \label{eq:def_Vij}
     V_{ij}= C_{ij}\frac{1}{4f_\pi^2}(p^0+p^{\prime0})\, ,
\end{eqnarray}
where $f_\pi=93$ MeV, and $p^0, p^{\prime0}$ are the energies of the incoming and outgoing mesons in the 
$PB\to P^\prime B^\prime$, etc. transitions. The $C_{ij}$ coefficients  for $I=0$ are easily evaluated and coincide with those obtained 
in~\cite{Xiao:2019gjd} for the dominant terms. We write the results in Tables~\ref{coeff_PB12}-\ref{coeff_VB} for the $PB(1/2^+)$ and $VB(1/2^+)$, 
respectively. The coefficient $\lambda$ corresponds to the exchange of $D^*$ or $D^*_s$, which is suppressed by approximately 
the ratio $(m_V/m_{D^*})^2$ (with $m_V=800$ MeV for the light vectors). As in \cite{Mizutani:2006vq} we take the value $\lambda=0.25$.
For the sectors $PB(3/2^+)$ and $VB(3/2^+)$, we have only one channel for each of them, $\bar{D}\Xi_c^*\to \bar{D}\Xi_c^*$ and 
$\bar{D}^*\Xi_c^*\to \bar{D}^*\Xi_c^*$, respectively, with the same value of $-1$ for the coefficient $C_{11}$.

As mentioned above, the dominant terms, this is to say those which remain finite in the heavy quark limit, compiled in 
Tables~\ref{coeff_PB12} and \ref{coeff_VB}, agree with those found in Ref.~\cite{Xiao:2019gjd}. In the later work, the subdominant 
terms, driven by the exchanged of a heavy meson, were fixed from the predictions of the LHG model for the $PB(1/2^+)$ block, 
and used, assuming HQSS, to all the rest of channels. This is inexact because the HQSS matrices derived in~\cite{Xiao:2019gjd} 
(Eqs.~(5), (6) and (7)) are only correct in the strict heavy quark limit. Indeed, there exist corrections when the interaction is due to the 
exchanges of heavy mesons which break the invariance under spin rotations of the heavy quarks. Here we evaluate this HQSS
terms within the LHG scheme.

%%%%%%%%%%%%-Table
\begin{table}[tbp]   %tbp H
%\footnotesize
\centering
 \caption{Coefficients $C_{ij}$ for the PB sector with $J^P = \frac{1}{2}^{-}$ and $I=0$.}
 \label{coeff_PB12}
\setlength{\tabcolsep}{6.5pt}
\setstretch{1.2}
\begin{tabular}{l|cccc}
\hline
\hline  %%$J^P = \frac{1}{2}^{+}$
          ~            &$\eta_c\Lambda\;$    & $\bar{D}_s\Lambda_c$       & $\bar{D}\Xi_c$                & $\bar{D}\Xi_c^\prime$ \\
\hline
  $\eta_c\Lambda$       & $0$    & $-\frac{1}{\sqrt{3}}\lambda$  & $\frac{1}{\sqrt{6}}\lambda$  & $-\frac{1}{\sqrt{2}}\lambda$   \\
  $\bar{D}_s\Lambda_c$  &       & $0$                                        & $-\sqrt{2}$                            & $0$                    \\
  $\bar{D}\Xi_c$                &      &                                                & $-1$                                      & $0$                    \\
  $\bar{D}\Xi_c^\prime$   &       &                                               &                                               & $-1$                   \\
\hline
\hline
\end{tabular}
\vspace{-0.5cm}
\end{table}
%%%%%%%%%%%%-Table

%%%%%%%%%%%%-Table
\begin{table}[tbp]   %tbp H
%\footnotesize
\centering
 \caption{Coefficients $C_{ij}$ for the VB sector with $J^P = \frac{1}{2}^{-}, \frac{3}{2}^{-}$ and $I=0$.}
 \label{coeff_VB}
\setlength{\tabcolsep}{6.5pt}
\setstretch{1.2}
\begin{tabular}{l|cccc}
\hline
\hline  %%$J^P = \frac{1}{2}^{+}$
          ~            &$J/\psi\Lambda\;$    & $\bar{D}^*_s\Lambda_c$       & $\bar{D}^*\Xi_c$          & $\bar{D}^*\Xi_c^\prime$ \\
\hline
  $J/\psi\Lambda$       & $0$    & $-\frac{1}{\sqrt{3}}\lambda$  & $\frac{1}{\sqrt{6}}\lambda$  & $-\frac{1}{\sqrt{2}}\lambda$   \\
  $\bar{D}^*_s\Lambda_c$  &       & $0$                                        & $-\sqrt{2}$                           & $0$                    \\
  $\bar{D}^*\Xi_c$                &      &                                                & $-1$                                      & $0$                    \\
  $\bar{D}^*\Xi_c^\prime$   &       &                                               &                                               & $-1$                   \\
\hline
\hline
\end{tabular}
\vspace{-0.5cm}
\end{table}
%%%%%%%%%%%%-Table

Here comes an important observation. As we see, in the matrices of sectors $1)$ and $3)$ in Tables~\ref{coeff_PB12}
and~\ref{coeff_VB}, respectively, the only non zero diagonal terms 
correspond to $\bar{D}\Xi_c$, $\bar{D}\Xi_c^\prime$ and $\bar{D}^*\Xi_c$, $\bar{D}^*\Xi_c^\prime$ with $C_{ii}=-1$. We should
not expect bound states for $\eta_c\Lambda$ and $\bar{D}_s\Lambda_c$ in $PB(1/2^+)$ sector or $J/\psi\Lambda$ and 
$\bar{D}^*_s\Lambda_c$ in $VB(1/2^+)$ sector. However, the transition $C_{ij}$ from $\bar{D}_s\Lambda_c$ to $\bar{D}\Xi_c$ or
$\bar{D}^*_s\Lambda_c$ to $\bar{D}^*\Xi_c$ is $-\sqrt2$, strong compared with the diagonal terms. 
The relevance of these two strongly coupled channels was already stressed in \cite{Chen:2022wkh},  however with a different interaction.

\begin{figure}[tbp]   %tbp H
  \centering
  \includegraphics[width=7cm]{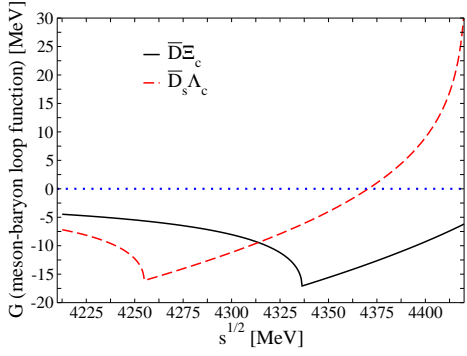}
  \caption{Re$G_l(\sqrt{s})$ for the $\bar{D}_s\Lambda_c$ and $\bar{D}\Xi_c$ channels using cut off regularization with
                $q_{max}=600$ MeV.
              }
  \label{fig1}
%\vspace{-0.2cm}
\end{figure}

\begin{figure}[tbp]   %tbp
  \centering
  \includegraphics[width=7cm]{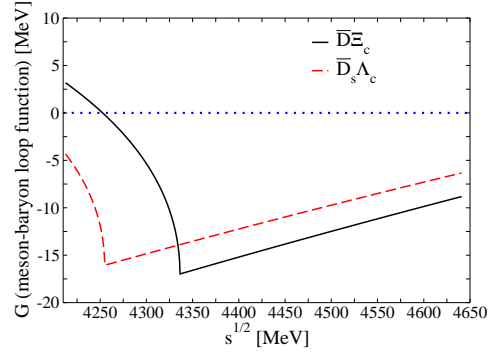}
  \caption{Re$G_l(\sqrt{s})$ for the $\bar{D}_s\Lambda_c$ and $\bar{D}\Xi_c$ channels using dimensional regularization.
                }
  \label{fig2}
\vspace{-0.5cm}
\end{figure}

To understand the effect of the coupled channels, let us recall that the molecular states are obtained by the solution of the Bethe-Salpeter 
equation in coupled channels
\begin{eqnarray}
        T = [1-VG]^{-1}V\,,
        \label{eq_T}
\end{eqnarray}
with $G$ the loop function of a meson-baryon intermediate state
\begin{eqnarray}
        G_l & =&  \int_{|{\bf q}|< q_{max}} \frac{d^3q}{(2\pi)^3}\frac{1}{2\omega_l({\bf q})}\frac{M_l}{E_l({\bf q})} \nonumber\\
                &\cdot&  \frac{1}{\sqrt{s}-\omega_l({\bf q}) - E_l({\bf q})+i\epsilon}\,,
          \label{eq_Gl}
\end{eqnarray}
where $\omega_l({\bf q})=\sqrt{m^2_l+{\bf q}^2}$ for the meson and $E_l=\sqrt{M^2_l+{\bf q}^2}$ for the baryon and $q_{max}$
is a regulator tied to the range of the interaction in momentum space \cite{Gamermann:2009uq,Song:2022yvz}. It is clear that this function is always negative below threshold $\sqrt{s}=M_{th}=m_l+M_l$. At $\sqrt{s}=M_{th}-\epsilon$, the real part Re$G$ of the 
loop function has infinite slope and for 
$\sqrt{s}>M_{th}$ it grows smoothly, but keeps negative in a long range of energies where one whishes to apply the theory. 
In the case of using dimensional regularization with non relativistic kinematics Re$G_l(\sqrt{s})$ for $\sqrt{s}>M_{th}$ is a 
constant, equal to the value of threshold, which we should take negative as it corresponds to the evaluation of Eq.~(\ref{eq_Gl})
at threshold with any cutoff \cite{Kaplan:1996xu}. In the present case and taking $q_{max}=600$ MeV suited to reproduce the results of 
Refs.~\cite{Xiao:2013yca,Xiao:2019gjd}, for $P_c$ and $P_{cs}$, where the dimensional regularization was used for the loops, the situation is as shown in 
Fig.~\ref{fig1}. In Fig.~\ref{fig2} we show the same $G$ functions in dimensional regularization, with relativistic kinematics, 
having the same value at the threshold as using the cutoff regularization with $q_{max}=600$ MeV.

%%%%%%%%%%%%%%%%%%%%%% Fig-1
%%%%%%%%%%%%%%%%%%%%%% Fig-1

%%%%%%%%%%%%%%%%%%%%%% Fig-2
%%%%%%%%%%%%%%%%%%%%%% Fig-2

As we can see, in the region of the $\bar{D}\Xi_c$ channel around $4338$ MeV, the real part of the two $G$ functions is negative and 
in Fig.~\ref{fig2} we observe that Re$G_{\bar{D}\Xi_c}$ can become positive below threshold as discussed in \cite{Wu:2010rv}, when 
dimensional regularization is used.

Let us take Eq.~(\ref{eq_T}) with the two relevant channels $\bar{D}\Xi_c$ and $\bar{D}_s\Lambda_c$, which we label now as 
$1$ and $2$, respectively, for the discussion that follows. We have a $V$ matrix of the type

\begin{equation}
\label{Vmatrix}
   \left(
      \begin{array}{cc}
             V_{11} & V_{12} \\
             V_{12} &  0       \\
      \end{array}
    \right)
\end{equation}
and Eq.~(\ref{eq_T})  is  immediately solved with the solution 
\begin{equation}
  T_{11}=\frac{V_{11}+V_{12}^2G_2}{1-(V_{11}+V_{12}^2G_2)G_1}\,.
   \label{eq_T11}
\end{equation}
This is also discussed in section~$6$ of~\cite{Aceti:2014ala} (see also \cite{Hyodo:2013nka}). Eq.~(\ref{eq_T11}) tells us that we can eliminate channel $2$ and 
work with only channel $1$ with an effective potential 
\begin{equation}
   V_{eff}=V_{11}+V_{12}^2G_2.
   \label{eq_Veff}
\end{equation}
If the channel $2$ had a higher threshold than channel $1$, Re$G_2$ would be always negative when looking for a bound state of 
channel $1$. In the opposite situation, as we have in Fig.~\ref{fig1}, Re$G_2$ can reach positive values, but, for $q_{max}=600$ MeV,
it is negative around the threshold of the $\bar{D}\Xi_c$ channel and below,  where the bound states are expected. 
One should note that the raise of Re$G_{\bar{D}_s\Lambda_c}$ (Re$G_2$) is tied to the relatively small cutoff. 
Indeed if $q_{max}=q_{on}$, where $q_{on}$ is the on shell 
momentum for the $\bar{D}_s\Lambda_c$ channel ($\sqrt{s}=\omega(q_{on})+E(q_{on})$), then Re$G\to +\infty$ as $\sqrt s$ increases, 
because there is only one branch to the left of $q_{max}$ in the evaluation of the principal value of the integral, and the branch 
above $q_{max}$ that would cancel the infinite is not allowed.
This is certainly not realistic and one should not use this $G$ function in that region, but the $G$ function is well behaved in the region where $\bar{D}\Xi_c$ bound states appear. To clarify further the situation we also plot
Re$G_{\bar{D}_s\Lambda_c}$ in dimensional regularization in Fig.~\ref{fig2} normalized to the same value at threshold, and 
we see that Re$G_{\bar{D}_s\Lambda_c}$ is always negative in the wide region of the plot. The consequence of this is that 
the effective potential has incremented $V_{11}$ (the $\bar{D}\Xi_c$ potential) by $V_{12}^2G_2$, 
which is negative in the region of relevance to us and, 
hence, the addition of the new $\bar{D}_s\Lambda_c$ channel has the effect of an extra attraction in the $\bar{D}\Xi_c$ channel. 
We have extended this discussion because the conclusion in Ref.~\cite{Chen:2022wkh} is opposite, and in that work the inclusion of the $\bar{D}_s\Lambda_c$ channel produces repulsion. While this is possible depending on the regularization procedure followed, 
our arguments strongly favor the general rule that the elimination of one channel reverts into extra attraction in the channel 
kept.

This has strong consequences in the comparison of the results that we obtain with those obtained for the $P_c$ states. Indeed, 
the diagonal channels corresponding to $\bar{D}\Xi_c$ and $\bar{D}_s\Lambda_c$ for the $P_c$ state in Ref.~\cite{Xiao:2013yca} are 
$\bar{D}\Sigma_c$ and $\bar{D}\Lambda_c$, but in \cite{Xiao:2013yca} the $C_{ij}$ coefficient for the transition between these latter channels (see Eq.~(29) of~\cite{Xiao:2013yca}) is zero. Thus, we should expect that the $\bar{D}\Xi_c$ state that we obtain here is more bound than the $\bar{D}\Sigma_c$ found in~\cite{Xiao:2013yca} at $4262$ MeV, which is bound by about $58$ MeV. Hence we should expect the binding of the 
$\bar{D}\Xi_c$ state reasonable bigger. This also means that we cannot by any means associate the new $P_{cs}(4338)$ state
to the $\bar{D}\Xi_c$ channel which has the mass $4336.3$ MeV, and it should correspond to a different channel. 
Even if we make the  $\bar{D}\Sigma_c$ less bound to correspond to the $P_c(4312)$, the  $\bar{D}\Xi_c$ state still will be 
bound by several tens of MeV and there is no way to make it correspond to the $P_{cs}(4338)$. On the other hand, 
$\bar{D}\Xi_c^\prime$ is the equivalent to $\bar{D}\Sigma_c$ (mixed symmetric spin wave functions) and one might expect
tens of MeV of bounding.

%After this discussion we present the results that we obtain and show the correspondence to the two $P_{cs}$ states.

%%%%%%%%%%%%%%%%%%%%%%%%%%%%
\section{Results}\label{sec-res} %%%========SEC-Sum

%%%%%%%%%%%%-Table
\begin{table*}[!]
%\footnotesize
\centering
\caption{Coupling constants and wave functions at the origin of all channels, corresponding to poles of the $T$ matrix in the 
            $PB(1/2^+)$ sector with $J^P=\frac12^-$. Masses and 
               half width in units of MeV, we stress in bold face the dominant channel.}
\label{coup_PB12}
\setlength{\tabcolsep}{8pt}
\setstretch{1.2}
\begin{tabular}{l|lcccc}
\hline 
\hline
Poles           &       ~            & $\eta_c\Lambda$    & $\bar{D}_s\Lambda_c$  & $\bar{D}\Xi_c$   & $\bar{D}\Xi_c^\prime$      \\
\hline
  \multirow{2}*{$4198.94+i0.11$} %\bm
                        &     $g_i$              & ${0.12-i0.00}$       & $3.01-i0.01$      & $\bm{4.85+i0.01}$    & $0.01-i0.03$          \\
            ~          &    $g_iG^{II}_i$   & ${-0.35+i1.01}$     & $-19.24+i0.05$  & $\bm{-20.35-i0.07}$ & $-0.03+i0.09$         \\
\hline
  \multirow{2}*{$4422.79 +i7.75$} %\bm
                        & $g_i$ 			& $0.71 -i0.08$ 	 & $0.05 +i0.00$ 	& ${-0.06+i0.04}$ 	& $\bm{2.79-i0.35}$     \\
            ~          & $g_iG^{II}_i$ 	& $1.14+i10.71$  	& $0.76+i1.82$ 	& ${-0.43-i1.53}$ 	& $\bm{-26.54+i0.64}$   \\
\hline
\hline
\end{tabular}
\end{table*}
%%%%%%%%%%%%-Table

%%%%%%%%%%%%-Table
\begin{table*}[!]
%\footnotesize
\centering
\caption{The same as in Table~\ref{coup_PB12}, but for the $VB(1/2^+)$ sector with $J^P=\frac12^-, \frac32^-$.}
\label{coup_VB12}
\setlength{\tabcolsep}{8pt}
\setstretch{1.2}
\begin{tabular}{l|lcccc}
\hline \hline
 Poles           &       ~    & $J/\psi\Lambda$    & $ \bar{D}^*_s\Lambda_c $  & $\bar{D}^*\Xi_c$   & $\bar{D}^*\Xi_c^\prime$      \\
\hline
  \multirow{2}*{$4337.98+i0.12$} %\bm
                        &     $g_i$              & ${0.11-i0.00}$       & $3.17-i0.01$      & $\bm{5.07+i0.01}$    & $0.00-i0.04$          \\
            ~          &    $g_iG^{II}_i$   & ${-0.13+i1.07}$     & $-18.57+i0.06$  & $\bm{-19.94-i0.07}$ & $-0.01+i0.10$         \\
\hline
  \multirow{2}*{$4565.73 +i15.58$} %\bm
                        & $g_i$ 			& $0.70 -i0.16$ 	 & $0.09 -i0.03$ 	& ${-0.10+i0.09}$ 	& $\bm{2.84-i0.72}$     \\
            ~          & $g_iG^{II}_i$ 	& $6.24+i21.04$  	& $2.07+i3.40$ 	& ${-1.37-i2.81}$ 	& $\bm{-26.31+i1.15}$   \\
\hline
\hline
\end{tabular}
\end{table*}
%%%%%%%%%%%%-Table

In Table~\ref{coup_PB12} we show the results that we obtain for the $PB(1/2^+)$ sector, showing the masses of the states, 
the couplings $g_i$ to each channel evaluated with $q_{max}=600$ MeV and the wave functions evaluated at the origin, $g_iG^{II}$ ($G^{II}$ is the loop function in the second Riemann sheet) \cite{Gamermann:2009uq}. In Table~\ref{coup_VB12} we show the equivalent 
results in the $VB(1/2^+)$ sector. In addition we have the single channel for $PB(3/2^+)$ and $VB(3/2^+)$,  with their poles and 
couplings  as %with the unit of MeV
\begin{eqnarray}
&&~~\;\bar{D}\Xi_c^*:~4488.52~{\rm MeV},~g=2.81,~gG^{II}=-26.47,\nonumber \\
&&~~\quad\;\qquad         J^P=\frac32^-;  \label{res_Dxicstar} \\
&&~\bar{D}^*\Xi_c^*:~4627.48~{\rm MeV},~g=2.96,~gG^{II}=-25.93,\nonumber \\
&&~~\quad\;\qquad         J^P=\frac12^-, \frac32^-, \frac52^-. \label{res_Dstarxicstar}
\end{eqnarray}

%\begin{itemize} 
%  \item [1)] $\bar{D}\Xi_c^*$:~$4488.52$ MeV, $g=2.81$, $gG^{II}=-26.47$, $J^P=\frac32^-$;
%  \item [2)] $\bar{D}^*\Xi_c^*$:~$4627.48$ MeV, $g=2.96$, $gG^{II}=-25.93$, $J^P=\frac12^-, \frac32^-, \frac52^-$.
%\end{itemize}  \label{res_xicstar}

%%%%%%%%%%%%%%%%%%%%%% Fig-3
\begin{figure}[tbp]   %tbp H
  \centering
  \includegraphics[width=5cm]{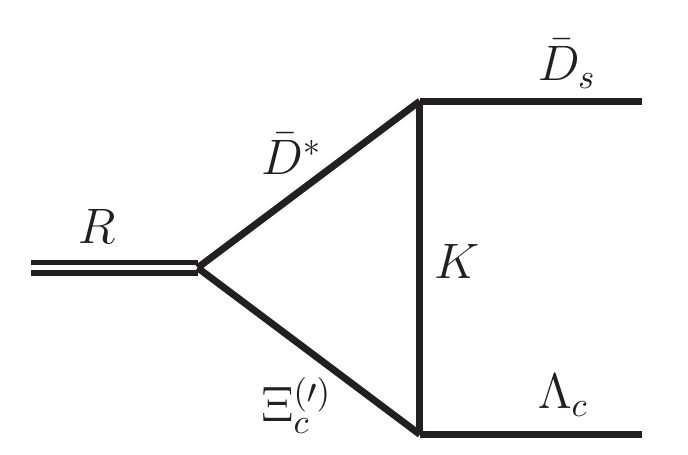}
  \caption{Loop diagram for the $\bar{D}^*\Xi_c^{(\prime)}$ state to decay to $\bar{D}_s\Lambda_c$.
                }
  \label{loop_Pcs}
%\vspace{-0.2cm}
\end{figure}
%%%%%%%%%%%%%%%%%%%%%% Fig-3

There is a caveat to be taken into account. The $\eta_c\Lambda$ channel has a threshold at $4100$ MeV, $326$ MeV below the 
state obtained in Table~\ref{coup_PB12} around $4423$ MeV. Channels far away in energy should not be relevant for the structure 
of the state obtained, except as being potential decay channels. This is why in Refs.~\cite{Dai:2020yfu,Wang:2019evy} 
a strategy was used, and proved successful,  which is to neglect the real part of the $G$ function of such channels and consider 
only their imaginary part. We use the same strategy here, which is further justified since for $q_{max}=600$ MeV we find a 
situation like the one of Fig.~\ref{fig1} for Re$G$ of the $\bar{D}_s\Lambda_c$ channel where Re$G_{\eta_c\Lambda}$ blows 
up at $\sqrt{s}\approx 4305$ MeV. We take here $G_{\eta_c\Lambda}\to i$Im$G_{\eta_c\Lambda}(M_{\rm inv})
=-i\frac{1}{4\pi}p_{\eta_c}\frac{M_\Lambda}{M_{\rm inv}}$, with $p_{\eta_c}$ the on shell momentum for $\eta_c$ in the decay 
of an object of energy $\sqrt s$ to $\eta_c\Lambda$.

Inspection of Tables~\ref{coup_PB12}-\ref{coup_VB12} shows that the state to be associated to the $P_{cs}(4338)$ is 
the state that couples mostly to $\bar{D}^*\Xi_c$ in Table~\ref{coup_VB12}. The mass coincides with the experimental value.
The width is very small, ($0.24$ MeV), with the state decaying to $J/\psi\Lambda$ which has a very small coupling. The
width would become a bit large if we consider the transition to the $PB(1/2^+)$ channels, essentially $\bar{D}_s\Lambda_c$ 
given the threshold masses. This requires the triangle loop of Fig.~\ref{loop_Pcs}. However, the dominant intermediate state 
$\bar{D}^*\Xi_c$ does not contribute because the $q\bar q \bm{\sigma}\cdot\bm{q}$ operator at the quark level, with  
the wave functions used in \cite{Xiao:2019gjd}, does not connect the $\Xi_c$ and $\Lambda_c$ states that go both with the spin $\chi_{MA}$
function. The  $\bar{D}^*\Xi_c^\prime$ channel can contribute, but we see in Table~\ref{coup_VB12} that the obtained coupling of 
the $4338$ MeV state to $\bar{D}^*\Xi_c^\prime$ is very small. We expect an extra very small width. Actually, the 
experimental width is $\Gamma_{P_{cs}(4338)}=7.0\pm1.2\pm1.3$ MeV, from where we should also 
exclude the effects of the experimental resolution. In this sense our predictions of a very small width should be considered as 
positive feature of the calculation. We should note that the state found also couples strongly to the $\bar{D}^*_s\Lambda_c$ 
component, although with weaker coupling than to $\bar{D}^*\Xi_c$. 

Furthermore, we should note that we get two degenerate states of $\frac12^-$ and $\frac32^-$ in this case. In practice, consideration of $\pi$ exchange to couple the $VB$ with $PB$ channels will break the degeneracy, if the two states are very close, this could justify 
the observed experimental results. 
Actually, this happened with the splitting of the $P_c$ state observed in~\cite{LHCb:2015yax} into two states observed 
in~\cite{LHCb:2019kea} with more statistics. Future measurements with higher resolution could tell us more about this suggestion.

Then we have to see a candidate for the $P_{cs}(4459)$. Our natural choice is the $PB(1/2^+)$ state obtained at $4423$ MeV
in Table~\ref{coup_PB12}, which couples mostly to $\bar{D}\Xi_c^\prime$. %The discrepancy of about $30$ MeV can be 
%considered within the uncertainties of the model. We have observed that small changes in $q_{max}$ change the mass of this 
%state much more than one at $4338$ MeV, and one could get a closer agreement. In favor of this association is the fact 
%that we obtain now a width of $28$ MeV, which is comparable with the experimental one  
%$\Gamma_{P_{cs}(4459)}=17.3\pm6.5^{+8.0}_{-5.7}$ MeV.
In favor of this association we should mention that we find a width of $15$ MeV which compares favorably with the 
experimental one of $\Gamma_{P_{cs}(4459)}=17.3\pm6.5^{+8.0}_{-5.7}$ MeV.

This reasonable association to the observed states still leaves some states unobserved. 
The one we predict at $4198.94$ MeV can decay to $\eta_c\Lambda$ according to Table~\ref{coup_PB12} and we should 
expect a small width, as shown by the calculation, given the small coupling of this 
state to $\eta_c\Lambda$.  The other states coupling mostly to $\bar{D}^*\Xi_c^\prime$, $\bar{D}\Xi_c^*$ and 
$\bar{D}^*\Xi_c^*$ appear at energies around $4500$ MeV and above. 
The $\bar{D}^*\Xi_c^\prime$ state, shown in Table~\ref{coup_VB12}, appears degenerate in our approach with 
$J^P=\frac12^-, \frac32^-$, while the $\bar{D}\Xi_c^*$ and $\bar{D}^*\Xi_c^*$, as shown in 
Eqs.~(\ref{res_Dxicstar})-(\ref{res_Dstarxicstar}), 
appear with $J^P=\frac32^-$ and $J^P=\frac12^-, \frac32^-, \frac52^-$, respectively. In total, and counting the different spins, 
we have $10$ states. They also show up at high energy in the molecular 
papers mentioned above and we are hopeful that it is a question of time that they are observed.

%%%%%%%%%%%%%%%%%%%%%%%%%%%%
\section{Further considerations}\label{sec-fur}%%%========SEC-further

\begin{figure}[tbp]   %tbp H
  \centering
  \includegraphics[width=7cm]{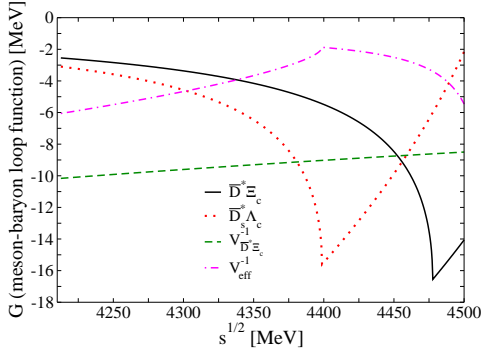}
  \caption{Re$G$ for the $\bar{D}^*\Xi_c$ and $\bar{D}^*_s\Lambda_c$ channels. The dashed line corresponds to 
                $V^{-1}_{\bar{D}^*\Xi_c,\bar{D}^*\Xi_c}$, the dashed dotted line to $({\rm Re}V_{eff})^{-1}$ (Eq.~(\ref{eq9})).
                }
  \label{fig4}
%\vspace{-0.2cm}
\end{figure}

Associating the state at $4338$ MeV to the $\bar{D}^*\Xi_c$ implies a binding of this channel by $140$ MeV. This is certainly 
bigger than the binding of the $\bar{D}\Sigma_c$ with respect to the $P_c(4312)$ of $13$ MeV. As we have already discussed,  
we expect a more binding here because of the effect of the coupled channel $\bar{D}^*_s\Lambda_c$. To make it more explicit 
we show in Fig.~\ref{fig4} how the state appears in the case of the $\bar{D}^*\Xi_c$ and $\bar{D}^*_s\Lambda_c$ coupled channels.

%%%%%%%%%%%%%%%%%%%%%% Fig-3
%%%%%%%%%%%%%%%%%%%%%% Fig-3

Should we have only one channel, $\bar{D}^*\Xi_c$, the pole appears when  
$V^{-1}_{\bar{D}^*\Xi_c,\bar{D}^*\Xi_c}=G_{\bar{D}^*\Xi_c}$. This produces a binding of about $20$ MeV, of the order of magnitude 
of the one of $\bar{D}\Sigma_c$ in the $P_c(4312)$ state. But, when we consider the effect of the $\bar{D}^*_s\Lambda_c$ 
coupled channel, as shown in Eq.~(\ref{eq_Veff}), this gives rise to a $\bar{D}^*\Xi_c$ effective potential
\begin{eqnarray}
  V_{eff}(\bar{D}^*\Xi_c)=V_{\bar{D}^*\Xi_c,\bar{D}^*\Xi_c}+
                                  V^2_{\bar{D}^*\Xi_c,\bar{D}^*_s\Lambda_c}\cdot G_{\bar{D}^*_s\Lambda_c}.
   \label{eq9}
\end{eqnarray}

We observe that now $V_{eff}^{-1}$ cuts Re$G_{\bar{D}^*\Xi_c}$ at the energy of $4338$ MeV as found in Table~\ref{coup_VB12}, 
with a binding much bigger than for the original single channel. We plot $({\rm Re}V_{eff})^{-1}$, but below the $\bar{D}^*_s\Lambda_c$ 
threshold where the pole appears $V_{eff}$ is real. It is logical to obtain this result. Indeed one can see that the additional potential 
in $V_{eff}$ is as large as $V_{\bar{D}^*\Xi_c,\bar{D}^*\Xi_c}$ which makes 
$V_{eff}^{-1}\approx\frac12(V_{\bar{D}^*\Xi_c,\bar{D}^*\Xi_c})^{-1}$ and this cuts Re$G_{\bar{D}^*\Xi_c}$ at much lower energies.
Should Re$G$ be a linear function, we would obtain a larger binding, but the slope of Re$G$ below the threshold is infinite and then 
gradually it becomes smaller in modulus such that a double potential means in practice a much bigger binding than 
we would obtain with a linear $G$ function, as one can appreciate in the figure.

On the other hand, the $\bar{D}\Xi_c^\prime$ channel does not benefit from the coupled channel effect since it only couples and 
weakly to the $\eta_c\Lambda$ channel, so one must rely in this case on the diagonal $\bar{D}\Xi_c^\prime$ interaction. Thus, 
with respect to the $\bar{D}\Xi_c^\prime$ threshold, the channel is only bound by about $22$ MeV, in line with the single potential 
binding of the $\bar{D}^*\Xi_c$ state. Yet, the experimental state $P_{cs}(4459)$ is about $13$ MeV above the 
$\bar{D}\Xi_c^\prime$ threshold, although with experimental errors it is closer. It is not easy for us to obtain a state above 
threshold, although the energy dependence of the potential allows it. However, should we admit a slightly different cutoff in the 
$PB$ than the $VB$ threshold and using $q_{max}=500$ MeV in the $PB$ sector, one obtains a mass of $4438$ MeV for the
state, closer to the experimental value, and one would obtain the result ever closer with small value of $q_{max}$ as used to 
obtain the $T_{cc}$ state~\cite{Feijoo:2021ppq} of the order of $450$ MeV. Admitting that a more complicated picture could emerge for 
this state, yet, we find it reasonable to accept that the channel  $\bar{D}\Xi_c^\prime$ with a threshold around the nominal mass of 
the $P_{cs}(4459)$ has much to do with that observed state. Incidentally, the state that appeared before at $4200$ MeV appears now 
at $4242$ MeV when $q_{max}=500$ MeV is used.
Hence, that state is more sensitive to the cutoff and one must also accept larger uncertainties in that prediction.
It will also be interesting to pay attention to future runs with more statistics for the experiment reporting that state, since other possible 
structures are hinted in the present spectrum.

%%%%%%%%%%%%%%%%%%%%%%%%%%%%
\section{Conclusions}\label{sec-conc}%%%========SEC-Conclu

%To sum up,
We have studied the interaction in coupled channels of pseudoscalar mesons and vector mesons with baryons of 
$J^P=1/2^+$ and $3/2^+$, evaluating the scattering matrix $T$, and looking for poles in the second Riemann sheet, from 
where we obtain the couplings looking at the residue of $T_{ij}$ at the pole. We obtain $10$ states coupling mostly to
$\bar{D}\Xi_c$, $\bar{D}^*\Xi_c$, $\bar{D}\Xi_c^\prime$, $\bar{D}^*\Xi_c^\prime$, $\bar{D}\Xi_c^*$ and $\bar{D}^*\Xi_c^*$.
We associated the $P_{cs}(4338)$ state to the one obtained with the same energy, coupling mostly to $\bar{D}^*\Xi_c$.
The $P_{cs}(4459)$ was associated to the state that couples mostly to $\bar{D}\Xi_c^\prime$ and the other states, 
one below the $J/\psi\Lambda$ threshold, which cannot be observed in this channel, and the other states appearing at 
higher energies, which are still unobserved.

The key point leading to this association was the realization that the $\bar{D}_s\Lambda_c$ and $\bar{D}^*_s\Lambda_c$ channels
play an important role given the strong transition potential from these channels to the $\bar{D}\Xi_c$ and $\bar{D}^*\Xi_c$ respectively, 
which has no counterpart in the corresponding $\bar{D}\Sigma_c$, $\bar{D}\Lambda_c$ and $\bar{D}^*\Sigma_c$, 
$\bar{D}^*\Lambda_c$ channels that appear in the evaluation of the $P_c$ states. Indeed, in this latter case the transition 
potentials between $\bar{D}\Sigma_c$ and $\bar{D}\Lambda_c$, or $\bar{D}^*\Sigma_c$ and $\bar{D}^*\Lambda_c$ are 
zero. We showed that this transition reverted into extra attraction. 
Hence, there is more attraction in the $\bar{D}\Xi_c$ or $\bar{D}^*\Xi_c$ channels than in the $\bar{D}\Sigma_c$ or 
$\bar{D}^*\Sigma_c$ ones and hence we should expect more binding in $\bar{D}\Xi_c$ and $\bar{D}^*\Xi_c$ than was found 
for $\bar{D}\Sigma_c$ and $\bar{D}^*\Sigma_c$ which produced the observed $P_c$ pentaquarks. 
It is then hard to associate the $P_{cs}(4338)$ directly to the $\bar{D}\Xi_c$ channel, which has a threshold mass at the 
same energy, and still be consistent with the commonly accepted molecular picture for the $P_c$ states.

In view of that, the natural picture that emerges from the present study is that the $P_{cs}(4338)$ state must be associated to 
the state coupling mostly to $\bar{D}^*\Xi_c$, while the $P_{cs}(4459)$ should be associated with the state coupling mostly 
to $\bar{D}\Xi_c^\prime$. The state coupling mostly to $\bar{D}\Xi_c$ is predicted below the $J/\psi\Lambda$ threshold 
and could be observed in $\eta_c\Lambda$. The other states appear at higher energies. These findings should stimulate further 
experimental research. They are also predicted in other molecular pictures at different energies and their observation would 
be most helpful to shed light on present theoretical approaches.

%%%%%%%%%%%%%%%%%%%%%%%%%%%%
\begin{acknowledgments}
This work was supported by the Spanish Ministerio de Economia y Competitividad (MINECO) and European FEDER funds 
under Contracts No. PID2020-112777GB-I00, and by Generalitat Valenciana under contract PROMETEO/2020/023.
This project has received funding from the European Union Horizon 2020 research and innovation programme under the 
program H2020-INFRAIA-2018-1, grant agreement No. 824093 of the STRONG-2020 project. The work of A. F. was partially supported by the Generalitat Valenciana and European Social Fund APOSTD-2021-112. 
This work of W. F. W. was supported in part by the National Natural Science Foundation of China under Grant No.~12147215. This work is also partly supported by the Fundamental Research Funds  for  the  Central  Universities  (JJW),  and  NSFC  under  Grant  No.   12070131001  (CRC110  cofunded  by  DFGand NSFC), Grant No.   11835015,  No.   12047503,  and by the Chinese Academy of Sciences (CAS) under Grants No.XDB34030000  (BSZ).

\end{acknowledgments}

%%%%%%%%%%%%%%%%%%%%%%%%%%%%

%=============== Refs ===============%  

%\bibliography{main}

\end{document}